\documentclass{acm_proc_article-sp}
\begin{document}

\title{Evaluation of Coordination Techniques in Synchronous Collaborative Information Retrieval}

\author{
%\alignauthor
Colum Foley and Alan F. Smeaton\\
       \affaddr{Centre for Digital Video Processing and Adaptive Information Cluster}\\
       \affaddr{Dublin City University}\\
       \affaddr{Glasnevin, Dublin 9, IRELAND}\\
       \email{colum.foley@computing.dcu.ie}
}
\maketitle
\begin{abstract}
Traditional Information Retrieval (IR) research has focussed on a single user interaction modality, where a user searches to satisfy an information need.
Recent advances in web technologies and computer hardware have enabled multiple users to collaborate on many computer-supported tasks, therefore there is an increasing opportunity to support two or more users searching together at the same time in order to satisfy a shared information need, which we refer to as \textit{Synchronous Collaborative Information Retrieval} (SCIR). SCIR systems represent a significant paradigmatic shift from traditional IR systems. In order to support effective SCIR, new techniques are required to coordinate users' activities. In addition, the novel domain of SCIR presents challenges for effective evaluations of these systems. In this paper we will propose an effective and re-usable evaluation methodology based on simulating users searching together. We will outline how we have used this evaluation in empirical studies of the effects of different  \textit{division of labour} and \textit{sharing of knowledge} techniques for SCIR.
\end{abstract}



 % NOT required for Proceedings

\section{Introduction}

 The purpose of an information retrieval (IR) system is to satisfy a user's information need. Traditionally, IR research has focussed on a single user interaction model.

\textit{Collaborative Information Retrieval} is a phrase that has been used to refer to many different technologies which support collaboration in the IR process. Much of the early work in collaborative information retrieval has been concerned with \textit{asynchronous}, remote collaboration. Collaborative filtering systems have been developed which attempt to reuse users' interactions with information objects in order to recommend them to others~\cite{goldberg92using}, collaborative re-ranking systems attempt to promote items of interest to a community of likeminded users~\cite{smyth05exploiting}, and collaborative footprinting systems record the paths of users through an information space so that others may follow~\cite{ahn05investigating}. Asynchronous collaborative information retrieval supports a passive, implicit form of collaboration where the focus is to improve the search process for \textit{an individual}.

Synchronous collaborative information retrieval (SCIR) systems represent a significant paradigmatic shift in information retrieval systems from an individual focus to a \textit{group} focus. SCIR systems are concerned with the realtime, explicit, collaboration which occurs when multiple users search together to satisfy a shared information need. This collaboration can take place either with the users working remotely, or, in a co-located setting. These systems have gained in popularity and now with the ever-growing popularity of the social web, and the development of new collaborative computer interfaces, there is a real opportunity to enable support for explicit, synchronous collaborative information retrieval.

%IR

%ASCIR

%SCIR
\section{Synchronous Collaborative Information Retrieval}
Early examples of SCIR systems include GroupWeb \cite{greenberg96groupweb} and W4 browser \cite{gianoutsos96collaborative}. The focus of these early SCIR systems were in increasing \textit{awareness} across collaborating users during a synchronised search, and this was achieved through various cues such as chat facilities, which users could use to communicate with each other, shared whiteboards, for realtime brainstorming, and bookmarking tools, where users could save documents of interest and bring them to the attention of the group. Although these systems allowed for a more engaging, collaborative search experience, providing awareness tools alone does not create effective SCIR. The benefit of allowing multiple users to search together in order to satisfy a shared information need is that it can allow for a \textit{division of labour} and a \textit{sharing of knowledge} across a collaborating group~\cite{zeballos98tools,foley06synchronous}. The awareness cues provided in early SCIR systems could allow users to coordinate their activities in order to achieve both a division of labour and a sharing of knowledge. For example, users could use a chat facility to divide the search task, e.g. ``You search for information on \textit{X} and I'll search for information on \textit{Y}", and the shared bookmark facility could enable a sharing of knowledge, as users can see the documents found by others. However, as noted by~\cite{adcock07fxpal}, requiring users to coordinate activities may become troublesome as it requires ``too much cognitive load to reconcile and integrate one's own activities with the opinions and actions of teammates''.

%Highlight Gap

Recently we have seen work which attempts to provide system-mediated coordination of users' actions in a collaborative search. In particular, the ``Cerchiamo" system of \cite{adcock07fxpal} was  a system for co-located video search which assigned co-searchers complementary roles and coordinated their activities by directing the group towards unexplored areas of the collection, the ``SearchTogether" system by \cite{morris07searchtogether} allowed users to divide the results of a search query across group members. Both of these systems represent ``first steps" towards effective system-mediated coordination of an SCIR search, however there is much still to explore.

%\section{Simulations in Synchronous Collaborative Information Retrieval}

\section{Evaluating SCIR}
\label{Section:Simulations}
In order to allow for the rapid evaluation of system-mediated techniques for synchronous collaborative information retrieval, novel evaluation methodologies are required. In this section we will outline a methodology which we have developed and which is based upon building simulations of two users searching together with an SCIR system.

\subsection{Simulations of SCIR}
%Simulations in IR
Simulations are used in information retrieval in an attempt to model a user's interactions with an IR system. A simulated user's interactions with a system can be controlled by using a parameterised user model and these models can vary in complexity based on the systems they are evaluating and the interactions they are attempting to simulate. Simulations are an attempt to bridge the gap of realism in information retrieval experimentation, between fully automatic experiments, where the user is taken out of the loop completely, and fully interactive experiments, where real users interact with an IR system.

%Simulations in SCIR

Previous IR experiments that have used user simulations have focussed on a single user's interactions with an IR system. In our work we are attempting to simulate a synchronous collaborative information retrieval environment, a dynamic, collaborative simulation. We will simulate a search involving \textit{two} collaborating users. Recent studies on the collaborative nature of search have shown how the majority of collaborative search sessions involve a collaborating group of two users \cite{morris07searchtogether} and therefore we believe that this group size is the most appropriate to model, though our proposed techniques could scale to larger group sizes.

When two or more users come together to search in an SCIR environment, there are several ways in which the collaborative search could be initiated. For example, users may each decide to formulate their own search query, or users may decide on a shared, group query. In either case, users are returned a set of documents to examine. As the search task proceeds, each user can examine their ranked list and may decide to view documents that seem relevant to the search task. Over the course of an SCIR search, users may read many documents related to the search task. If users find documents relevant to the search, they may decide to bookmark these documents in order to bring them to the attention of the group. Users may also decide to reformulate their search query during the search.

 \begin{centering}
\begin{figure}[h]
\centering \epsfig{file=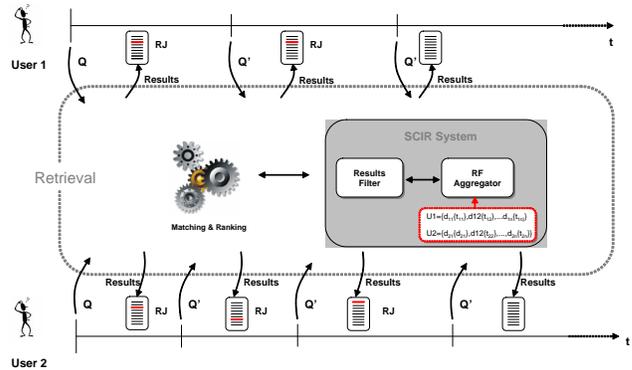, width=0.5\textwidth}
\caption{Conceptual overview of an SCIR session involving two users}
\label{Figure:Interaction_Diagram_2}
\end{figure}
\end{centering}

\noindent
Figure \ref{Figure:Interaction_Diagram_2} presents a conceptual overview of two users collaborating using the SCIR system described thus far.  Referring to this figure, the data required to populate our SCIR simulations consists of:
\begin{itemize}
\item Queries (Q) --  an SCIR search begins with a query entered by the users and users may decide to reformulate their queries during the search.
\item Series of relevance judgments (RJ) -- these are explicit
indications of relevance made by a user on a particular document. State-of-the-art SCIR systems allow for relevance judgments to be made in the form of bookmarks.
\item Timing information -- this represents the time, in seconds, relative to the
start of the search session, at which events such as relevance judgments and intermediate query reformulations are made. This information is used to order events in an SCIR simulation.
 \end{itemize}

Having outlined the requirements for an SCIR simulation, we will now describe how we populated our simulations using data from previous TREC interactive experiments.

\subsubsection{Populating Simulations with TREC Data}

The purpose of the TREC (Text REtreival Conference) interactive search is for a searcher to locate documents of relevance to a stated information need (a search ``topic") using a search engine and to save them \cite{nist97trec}. Each participating group that submitted results for evaluation in TREC 6 to TREC 8 was required to also include \textit{rich format data} with their submission.  This data consisted of transcripts of a searcher's significant events during a search and their timing information.

\begin{figure}
\epsfig{file=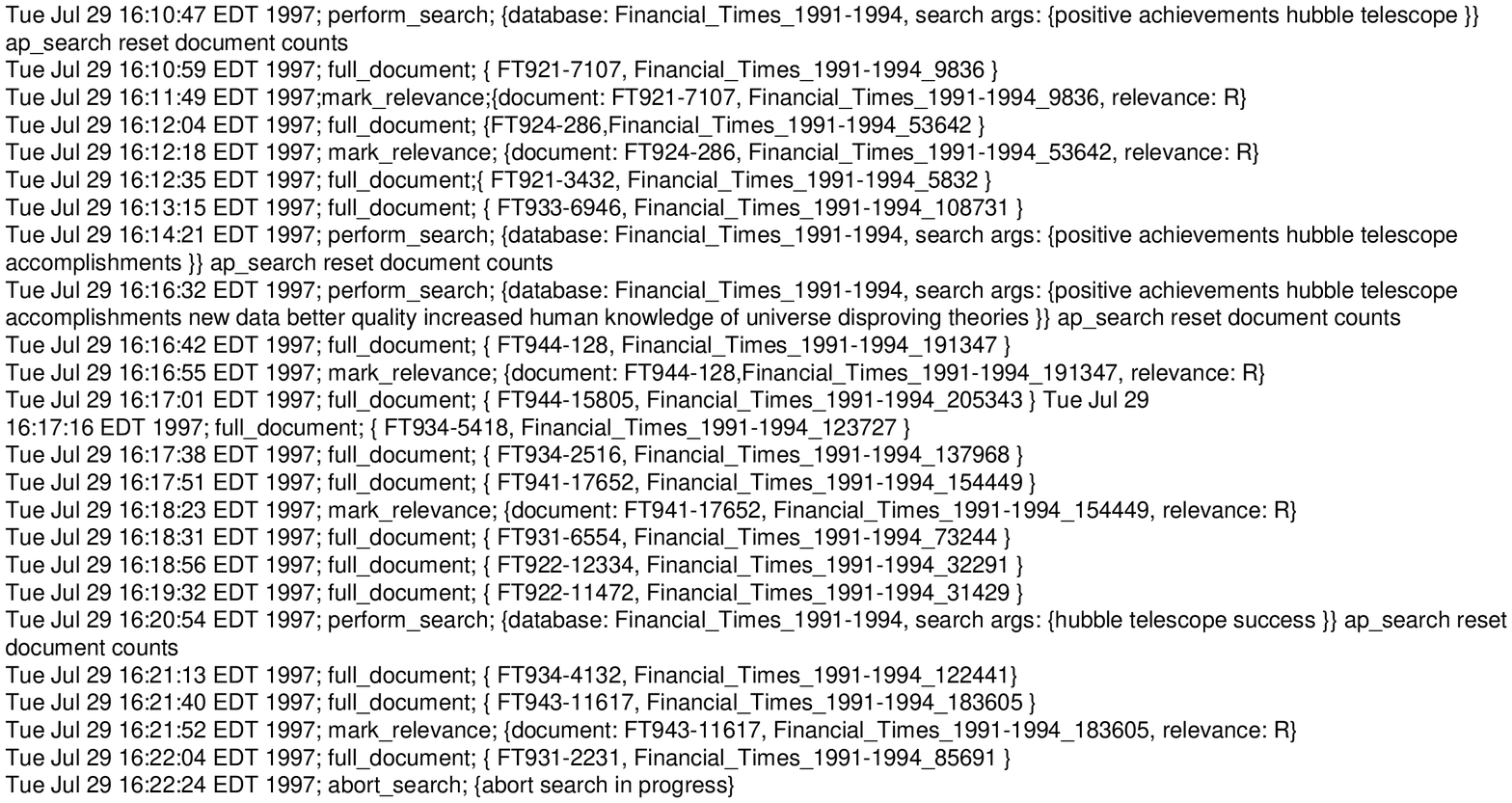, width=0.5\textwidth}
\caption{UMASS TREC 6 rich format data}
\label{Figure:rich_text_example}
\end{figure}

Figure \ref{Figure:rich_text_example} shows a sample rich format transcript from a user who completed topic 303i, entitled ``Hubble Telescope Achievements'', as part of the University of Massachusetts TREC 6 submission. Here we can identify queries (\textit{perform\_search}), relevance judgments (\textit{mark\_relevance}), and timing information (\textit{16:22:24}).  We can see that the user began their search by entering the query \textit{``positive achievements hubble telescope''}. After 62 seconds the user made a relevance judgment on document
\textit{FT921-7107} and the user provided a further 4 relevance judgments, and 3 query reformulations, until the search session finished after 697 seconds. As part of each participating group's TREC experiments several users would have completed the same search topic. Originally, these users would have performed these topic searches independently, for our simulations we model these users searching together synchronously in groups of two. In order to simulate these users searching at the same time, we synchronise their session start times by aligning the times for their initial query. We then arrange the significant events of the two users in time-order using the timing offsets from each user's data.

\begin{centering}
\begin{figure}[h]
\centering \epsfig{file=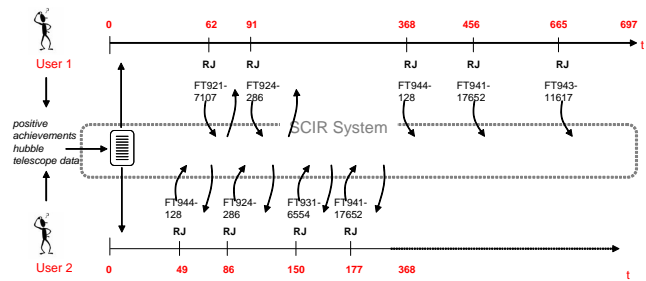, width=0.5\textwidth}
\caption{SCIR simulation using TREC rich format data}
\label{Figure:eventlist2}
\end{figure}
\end{centering}

Figure~\ref{Figure:eventlist2} shows an example of how this TREC data can be used to simulate an SCIR session involving two users. In this example, user 1 represents the user whose data is shown in Figure~\ref{Figure:rich_text_example}, and user 2 is another user who completed this search topic as part of the original UMASS submission. Here we can see that the search begins with a single group query  ``positive achievements hubble telescope data''. In this example, we do not show the intermediate query formulations and instead just show the relevance judgments made by users during the search.

By extracting rich format data associated with different users' interactions on a search topic, we can acquire multiple heterogenous simulations, where the data populating our simulations is from real users searching to satisfy the same information need on a standardised corpus.

Before we can finalise our simulations we need to resolve some outstanding issues relating to the static nature of the rich format data used to populate the simulations and the dynamic nature of the SCIR environment we are attempting to simulate. In particular, as we are applying the transcripts of a real user's interactions with a particular search system, to a dynamic, collaborative SCIR simulation, in order for the simulations to remain realistic, in the simulations we need to replace the actual documents saved by users during the original TREC runs (e.g. ``FT921-7101" from Figure~\ref{Figure:eventlist2}) with documents contained on the ranked lists that are presented to the simulated searchers. Otherwise, if we chose to keep the relevance judgments made by simulated users the same as the documents saved by the real users, we would be assuming that users would save the same document regardless of what list was presented to them from an SCIR system, an obvious oversimplification. Instead, when simulating a user making a relevance judgment in our system, our solution is to simulate the user making a relevance judgment on the first \textit{relevant} document they encounter on their ranked list, where the relevance of the document is obtained from the relevance judgments for the topic (or \textit{qrels}) from TREC.

\subsection{Evaluation Metric for SCIR}

At each stage in an SCIR session, each collaborating user will have associated with him/her a ranked list of documents. In traditional, single-user, information retrieval the accuracy of ranked lists can be evaluated using standard IR measurements such as average precision (AP). In our work we are concerned with the performance of a \textit{group} of users and therefore we need to be able assign a score to the collaborating group at any particular point in the search process.

%AP
One potential method for generating this \textit{group score}, would be to evaluate the quality of each collaborating searcher's list using a standard IR measure like AP then average these values across group members to get the average score for the group. Unfortunately, this approach of generating a group score does not adequately measure the group's performance as no attempt is made to examine the contents of the users' ranked lists and, in particular, the amount of overlap between them. To illustrate this further, if two separate collaborating groups of users have the same associated group score, arrived at by averaging the AP of each group member's ranked list, but the members of the first group had ranked lists which contained many of the same documents, while the second group had ranked lists with a greater diversity of relevant documents, then the performance of the second collaborating group should be considered better than the first as, across the group, the total amount of relevant material found across collaborating users' lists is greater in the second group. By simply averaging each individual's AP scores, however, this information is lost.

What we need instead is a measure which captures the quality and diversity across collaborating users' ranked lists. We propose to measure the \textit{total number of unique relevant documents across user's ranked lists} at a certain cutoff and use this figure as our \textit{group score}. This performance measure will capture both the quality and diversity across collaborating users' ranked lists and in particular the parts of the list are of interest to the SCIR system designer. The cut-off value can be set at different ranked positions, e.g. top 10, 20, 30, to see the number of unique relevant documents found across users' lists at different positions in the ranking.

\section{Division of Labour and Sharing of Knowledge in SCIR}
\label{Section:Techniques}

In our work we are interested in exploring the effects of a system-mediated \textit{division of labour} and \textit{sharing of knowledge} on the performance of a group of users searching together. Division of labour enables each collaborating group member to explore a subset of a document collection by limiting the overlap of results across users in order to improve the effectiveness of the search. Sharing of knowledge enables collaborating users to benefit from the activities and discoveries of their collaborators.

In our evaluations we used the simulations described in the previous section to simulate two users searching together through a simple incremental relevance feedback SCIR system. We simulated two searchers deciding on an initial query with which to begin the collaborative search and then simulated each searcher providing relevance judgments, with each relevance judgment initiating a relevance feedback iteration thereby returning a new ranked list to the user. We then implemented various types of division of labour policies on the ranked lists returned to users. We also explored the effects of an automated sharing of knowledge through both \textit{collaborative} and \textit{complementary} relevance feedback processes. Due to space restrictions, we are unable to discuss the details of our experiments here, instead we will provide an overview of the work:
\begin{itemize}
\item Division of labour -- we have examined the effects of implementing several division of labour techniques, whereby the results returned to collaborating users are automatically filtered to ensure an effective division of the search task across users.
\item Sharing of knowledge -- a common feature of state-of-the-art SCIR systems is their use of a bookmarking facility, where users can save documents of relevance to the group. We have experimented with providing relevance feedback mechanisms for SCIR whereby these bookmarks can be incorporated into a relevance feedback process. We have extended the traditional relevance feedback mechanism to allow for the combination of multi-user relevance information in a \textit{collaborative relevance feedback} process and have evaluated its effects on SCIR.
\item Sharing of knowledge under imperfect relevance information -- In our work we have modelled SCIR environments in which users can make mistakes in their relevance judgments and have evaluated the effects of this on a collaborative relevance feedback process.
\item Authority weighting -- we have implemented techniques to limit the effects of poor relevance assessments on a collaborative relevance feedback process through attaching an authority weight to users' relevance assessments and using this weight in a user-biased collaborative relevance feedback process.
\item Complementary relevance feedback -- a complementary feedback process leverages each users' relevance judgments in an SCIR search in order to promote diversity across users' ranked lists by reformulating each user's query in such a way as to make it as diverse as possible from their search partners'.
\end{itemize}

Our results show that both a division of labour and a sharing of knowledge policy can improve the effectiveness of two users searching together through an SCIR system, with the largest improvement being achieved through a division of labour. Encouragingly, this means that our empirical evaluation of SCIR has demonstrated that system-mediated SCIR search is more effective than users searching independently.

\section*{Acknowledgments}
This work is supported by the Irish Research Council for Science Engineering and Technology, and by Science Foundation Ireland under grant 03/IN.3/I361.

\bibliographystyle{abbrv}

\end{document}